# Quantum phase slips in superconducting wires with weak inhomogeneities

Mihajlo Vanević and Yuli V. Nazarov
*Kavli Institute of Nanoscience, Delft University of Technology, 2628 CJ Delft, The Netherlands*
( Dated: February 13, 2012)

Quantum phase slips are traditionally considered in diffusive superconducting wires which are assumed homogeneous. We present a definite estimate for the amplitude of phase slips that occur at a weak inhomogeneity in the wire where local resistivity is slightly increased. We model such a weak link as a general coherent conductor and show that the amplitude is dominated by topological part of the action. We argue that such weak links occur naturally in apparently homogeneous wires and adjust the estimate to that case. The fabrication of an artificial weak link would localize phase slips and facilitate a better control of the phase-slip amplitude.

The phase-slip processes in superconducting wires and long Josephson-junction arrays remain an active research subject both experimentally and theoretically [1–4]. In the course of a phase slip, the superconducting order parameter fluctuates to zero at a point in the wire while the superconducting phase difference along the wire changes by $\pm 2\pi$. The *incoherent* phase slips provide a mechanism for superconducting wires to retain a finite resistance at temperatures below the superconducting transition. Phase-slip events are thermally activated at temperatures close to critical [5] and triggered by quantum fluctuations at low temperatures [6]. Progress in microfabrication has enabled production of superconducting wires with diameters of a few tens of nanometers in which incoherent quantum phase slips have been studied experimentally [7–10]. Recently, much attention has been paid to *coherent* phase slips [11–14]. It has been argued that a wire where coherent phase slips take place may be regarded as a new circuit element – the phase-slip junction [12] – which is a dual counterpart of the Josephson junction with superconducting phase difference replaced by charge. The phase-slip qubit [11] [see Fig. 1(b)] and other coherent devices [13] have been proposed. The novel functionality may be useful in realization of the fundamental current standard dual to the Josephson voltage standard [12].

The coherent phase-slips in a wire are characterized by a quantum amplitude $E_S$ rather than a rate of an event [1, 15]. The amplitude depends exponentially on the instanton action which is usually dominated by the phase-slip "core" $\mathcal{S}_{\rm core} = \alpha (G_Q R' \xi)^{-1}$ where $R'$ is a wire normal-state resistance per unit length, $\xi$ is the coherence length, and $G_Q \equiv e^2/\pi\hbar$ (hereafter $\hbar = 1$). The numerical factor $\alpha$ depends on the details of the "core" profile which are unknown. Therefore, the amplitude $E_S \propto e^{-\mathcal{S}_{\rm core}}$ is exponentially small for not very resistive wires and is difficult to predict for the specific experimental settings since even a small arbitrariness in $\alpha$ would amount to orders of magnitude ambiguity in $E_S$ [11].

In this Letter, we report on a definite estimate of $E_S$ [Eq. (1)] for a weak link in diffusive wire where resistivity is slightly and locally enhanced. We argue that such weak links occur naturally in apparently homogeneous wires and adjust the estimate to that case as well.

To justify the model, let us first note that much attention is paid experimentally to making the wires as homogeneous as possible [10]. Indeed, if the resistance of the wire is dominated by a single weak link, the device would be a Josephson junction which is the opposite of the phase-slip junction intended. However, a *weak* inhomogeneity, where the local resistivity of the wire is only *slightly* larger, will not spoil the phase-slip character of the junction. The condition for this is just that the resistance of the weak link is much smaller than the overall normal-state resistance of the wire. Such weak links occur naturally in apparently homogeneous wires. Owing to exponential dependence on resistivity, the phase slips will be localized at the weak links. Thus, making such a weak link artificially would provide a better control for $E_S$, since one knows where the phase slips occur.

This motivates us to consider a simple yet general model of a weak link where the link is described as a short (length much smaller than $\xi$) coherent conductor characterized by a set of spin-degenerate transmission eigenvalues $\{T_p\}$. We solve this model and obtain the accurate estimate for the amplitude

$$E_S \approx 2\Delta \sqrt{\sum_p T_p} \prod_p \sqrt{1-T_p} \tag{1}$$

under approximations specified in the text, where $\Delta$ is the superconducting order parameter in the wire. We show that the amplitude is dominated by the topological part of the action emerging from $\pm 2\pi$ phase winding in the phase-slip process. Finally, we use the known transmission distribution of a diffusive conductor and obtain an estimation of $E_S$ valid for homogeneous wires as well.

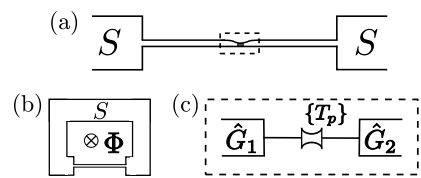

FIG. 1. (a) Superconducting diffusive wire with a weak link (dashed rectangle) connecting bulk superconducting electrodes. (b) Embedding the wire into a superconducting loop makes a phase-slip flux qubit [11]. (c) We model the weak link as a general coherent conductor characterized by a set of transmission eigenvalues $\{T_p\}$.

The system under consideration is depicted in Fig. 1. The weak link (or, "contact") is modelled as a general coherent conductor with conductance $G_c = G_Q \sum_p T_p$. The wire is much thinner than $\xi$ and is characterized by the length $L$ ($L \gg \xi$), normal-state resistance $R'$, and capacitance $C'$, where $'$ signifies that these quantities are defined per unit length. For a wire thickness in tens of nanometers range, the geometric inductance $\mathcal{L}'_g$ is negligible with respect to the kinetic inductance $\mathcal{L}'_k \equiv R'/\pi\Delta$. For concreteness, we consider the wire in a phase-slip qubit configuration [Fig. 1(b)]. This does not affect the evaluation of $E_S$.

Generally, the quantum dynamics of such systems is described by an imaginary-time action that is path-integrated over fluctuating superconducting order parameter $\Delta(\tau, x)$, where $x$ is the coordinate along the wire. Our model brings about drastic simplifications. The modulus of order parameter can be regarded as constant, its phase $\phi(\tau, x)$ being the only dynamical variable. The action comprises two terms, $\mathcal{S}[\phi] = \mathcal{S}_c[\phi] + \mathcal{S}_w[\phi]$, which describe the weak link and the wire, respectively. The action $\mathcal{S}_c$ for tunnel coupling was obtained in [16]. We generalize the result to generic coherent contact along the lines of Ref. [17]. The action reads

$$\mathcal{S}_c = -\frac{1}{2} \sum_p \operatorname{Tr} \ln\left(1 + \frac{T_p}{4}(\{\hat{G}_1, \hat{G}_2\} - 2)\right) \quad (2)$$

with $\hat{G}_j(\tau, \tau') = e^{i\phi_j(\tau)\hat{\tau}_3/2} \hat{G}_0(\tau - \tau') e^{-i\phi_j(\tau')\hat{\tau}_3/2}$. Here $\hat{G}_{1,2}$ are imaginary-time Green's functions in a wire on the left and right side of the weak link [cf. Fig. 1(c)], $\phi_{1,2}$ are the corresponding phases, $\hat{G}_0(\omega) = (\omega \hat{\tau}_3 + |\Delta|\hat{\tau}_1)/\sqrt{\omega^2 + |\Delta|^2}$ is the Green's function of a homogeneous superconductor, and $\hat{\tau}_i$ are the Pauli matrices in Nambu space. We see that this action depends on the phase difference $\phi(\tau) \equiv \phi_2(\tau) - \phi_1(\tau)$ only.

The resistance of the weak inhomogeneity in the wire is naturally much smaller than the total resistance of the wire, $R_c \ll LR'$ ($R_c \equiv G_c^{-1}$). The same pertains to inductance. Under these conditions, the minima of the action correspond to a well-defined fluxon states where the winding of the phase along the wire takes values $2\pi n$, $n$ being integer. The energies of the states are given by $E_n = (\Phi - n\Phi_0)^2/2L\mathcal{L}'_k$, where $\Phi$ is the flux penetrating the loop and $\Phi_0 = \pi/e$ is the flux quantum. Technically, it is convenient to ascribe the phase difference to the weak link and concentrate on $\Phi = \Phi_0/2$ where minima $n = 0, 1$ are degenerate. The phase-slip amplitude $E_S$ is then computed from analysis of instantons in $\phi(\tau)$ connecting these two energy-degenerate minima and equals to the energy splitting of the resulting qubit states [11].

The wire provides an electromagnetic environment for the phase propagation. In our situation, $\xi \partial_x \phi \ll 1$ and the effective environment is linear. Owing to this, the quadratic action $\mathcal{S}_w$ can be expressed in terms of $\phi(\tau)$ [18], $\mathcal{S}_w[\phi] = (8\pi^2 G_Q)^{-1} \int_0^\infty d\omega\, \omega\, Y(\omega)\, |\phi(\omega)|^2$, where $Y(\omega) = [\mathcal{L}'(\omega)/C']^{-1/2}[\tanh(\omega L_1/v_p) + \tanh(\omega L_2/v_p)]^{-1} - (L\mathcal{L}'_k \omega)^{-1}$, $L_1$ ($L_2$) is the length of the wire left (right) from the contact and $v_p(\omega) = 1/\sqrt{\mathcal{L}'(\omega)C'}$. Here $\mathcal{L}'(\omega)$ is the inductance in imaginary frequency obtained by analytic continuation of impedance. It accounts for the fact that the wire is inductive with $\mathcal{L}' = \mathcal{L}'_k$ at subgap energies $\omega \ll 2\Delta$, and resistive with $\mathcal{L}'(\omega) = R'/\omega$ at large energies $\omega \gg 2\Delta$. This completes theoretical description of the model. The instanton solution $\phi(\tau)$ minimizes $\mathcal{S}[\phi]$ satisfying $\phi(-\infty) = 0$ and $\phi(\infty) = 2\pi$.

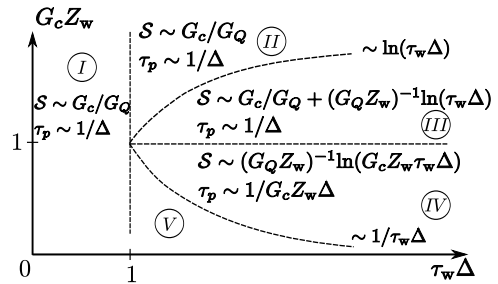

FIG. 2. The phase-slip regimes (see text) in parameter space $(G_c Z_w, \tau_w \Delta)$ where $Z_w$ is the wave impedance of the wire and $\tau_w$ is the characteristic time of charge propagation through the wire. We concentrate on the regions $I$ and $II$ where $E_S$ does not depend on the wire parameters.

We want to concentrate on the case when the estimation of $E_S$ does not depend on wire parameters. This is not always so and we need to discuss various regimes that may be realized in the system (Fig. 2). The relevant wire parameters are the wave impedance $Z_w = \sqrt{\mathcal{L}'_k/C'}$ and the characteristic charge propagation time $\tau_w$, which is estimated as either plasmon propagation time $L\sqrt{\mathcal{L}'_k C'}$ ($\tau_w \Delta \gg 1$, superconducting response) or $RC$-time $L^2 R' C'$ ($\tau_w \Delta \ll 1$, dissipative response). Let $\tau_p$ be the optimal instanton duration. The weak-link action can be then estimated as $\mathcal{S}_c \simeq (G_c/G_Q) \max(1, \tau_p \Delta)$. As to the wire action, it corresponds to the dissipative response $\mathcal{S}_w \simeq (G_Q Z_w)^{-1} \ln(\tau_w/\tau_p)$ if charge propagation does not reach wire ends for the time $\tau_p$ and to the capacitive response $\mathcal{S}_w \simeq LC'/G_Q \tau_p$ otherwise. The $\tau_p$ is found from minimizing $\mathcal{S} = \mathcal{S}_c + \mathcal{S}_w$ which gives rise to five regimes depicted in Fig. 2.

For "short" wires ($\tau_w \Delta \ll 1$) the action is dominated by the weak link and $\tau_p \simeq 1/\Delta$ (region $I$). For "long" wires ($\tau_w \Delta \gg 1$), we encounter the variety of regimes. At sufficiently large $G_c$, the above estimations still hold (region $II$). Upon decreasing $G_c$, the inductive wire response starts to dominate while $\tau_p \simeq 1/\Delta$ (region $III$). At $G_c \simeq Z_w$, the instanton duration $\tau_p$ increases. It is determined from the competition of inductive response of the weak link and inductive response of the wire (region $IV$). Upon further decrease of $G_c$, the $\tau_p$ matches $\tau_w$. Below this, the wire response is capacitive and $\tau_p$ is determined from the competition of inductive response of the weak link and the capacitive response of the wire (region $V$), very much like in traditional theory of macroscopic phase tunnelling [16]. We conclude that there is a large part of the parameter space (regions $I$, $II$) where instanton action is dominated by $\mathcal{S}_c$ and concentrate on the minimization of this part of the action.

For an arbitrary transmission set $\{T_p\}$ the analytical solution cannot be obtained, and we have treated the problem numerically [19]. However, the analysis of the numerical results permitted to formulate a good analytical approximation. To outline this, let us note that the action in Eq. (2) can be expressed in terms of the eigenvalues $\Lambda_n$ of a Hermitian operator $\hat{\Lambda} \equiv (\hat{G}_1 - \hat{G}_2)/2$,

$$\mathcal{S}_c[\phi] = -\frac{1}{2} \sum_{p,n} \ln(1 - T_p \Lambda_n^2). \quad (3)$$

One can deduce some properties of the eigenvalues that do not depend on details of the instanton profile $\phi_{\rm in}(\tau)$. First of all, $|\Lambda_n| \leq 1$. Importantly, there is a single eigenvalue precisely at $\Lambda = 1$. This is guaranteed by topological properties of $\hat{\Lambda}$ with respect to variations of $\phi_{\rm in}(\tau)$; similar discussion is provided in [20]. Generally, the number of these special eigenvalues is set by the winding number of $\phi(\tau)$, which is 1 in the case under consideration. All other eigenvalues come in pairs $\pm \Lambda$.

The special eigenvalue gives a topological contribution to the action

$$\mathcal{S}_{c1} = -\frac{1}{2} \sum_p \ln(1 - T_p) \quad (4)$$

which presents a lower bound for $\mathcal{S}_c$. This lower bound could have been realized if there was an instanton profile for which all non-special $\Lambda_n$ are zero. In the normal-metal case such instantons indeed exist and can even be found analytically [21]. This is not the case for superconducting action. However, the numerics prove that for the optimal instanton all non-special $\Lambda_n$ are small and the topological contribution gives an accurate estimation of the overall action. For instance, in the tunnel limit ($T_p \ll 1$) $\mathcal{S}_c = 0.528\,G_c/G_Q$ while the topological bound is $\mathcal{S}_{c1} = 0.5\,G_c/G_Q$. In all cases investigated, relative accuracy of topological approximation was better than 6%. Formally, the exponential dependence of $E_S$ could amplify even this small error by orders of magnitude; yet this does not happen for any $E_S$ of interest (see Fig. 3).

This gives us the value of the action. We also need to compute the prefactor. The prefactor is evaluated by the standard instanton techniques yielding $E_S = 2(\int d\tau \dot{\phi}_{\rm in}^2/2\pi)^{1/2}\,(D')^{-1/2}\,e^{-\mathcal{S}_{\rm in}}$. The ratio of determinants $D' = {\det}'(\delta^2 \mathcal{S}/\delta\phi^2|_{\rm in})/\det(\delta^2 \mathcal{S}/\delta\phi^2|_0)$ takes into account fluctuations with respect to instanton and trivial

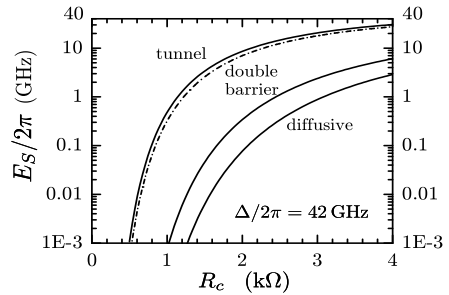

FIG. 3. Phase-slip amplitude $E_S$ for tunnel contact, double-barrier junction, and short diffusive bridge estimated using topological action $\mathcal{S}_{c1}$ (solid curves). The true $E_S$ with non-topological contribution taken into account is shown in the tunnel limit (dash-dotted curve).

trajectories; the prime $'$ denotes that the zero eigenvalue intrinsic to instanton is omitted in the numerator.

It is important to note that the high eigenvalues $h_n$ at $n \gg 1$ of $\delta^2 \mathcal{S}/\delta\phi^2$ are linear in $n$. This is related to the frequency dependence of the integral kernels in the action: for rapidly varying $\phi(\tau)$, the action reads $\mathcal{S}_c = (G_c/16\pi^2 G_Q) \int d\omega\,|\omega|\,|\phi(\omega)|^2$ (assuming $\omega \gg \Delta$). This implies logarithmic divergence of $\ln(D')$ at large energies. In principle, account of the wire capacitance might provide an upper cut-off needed. However, we find it more consistent to cancel the divergence by taking into account the renormalization of transmission eigenvalues.

Indeed, it is known that Coulomb interaction leads to energy-dependent renormalization of $T_p$ [22]. Under current-bias conditions, which is the case under consideration, the renormalization reads: $dT_p/d\ln E = T_p(1 - T_p)/\sum_p T_p$. Correcting the transmissions in $\mathcal{S}_{c1}$ with the above equation indeed cancels the divergence of $(D')^{-1/2}$. It implies that the $T_p$ in all formulas must be taken at $E \simeq \Delta$ rather than at unphysical high energy. The procedure is similar to common treatment of ultra-violet divergencies in the instanton determinant [23]. This brings us to Eq. (1). We stress that by virtue of instanton approximation, this relation is only valid for $E_S \ll \Delta$.

To make concrete predictions (Fig. 3), we need to specify type of the weak link. Using known transmission distributions [24] we find $\mathcal{S}_{c1} = \alpha\, G_c/G_Q$ with $\alpha = 1/2$, 1, $\pi^2/8$ for a tunnel junction, double tunnel junction, and diffusive weak link, respectively. The phase-slip amplitude $E_S$ for these types of weak links is shown in Fig. 3 for $T_c = 1.2\,{\rm K}$. For qubit applications, $E_S$ should be in the gigahertz range. In this range, $E_S$ at a given $R_c$ varies by two orders of magnitude depending on the type of the weak link. Dash-dotted curve for tunnel junction illustrates the accuracy of topological approximation.

Let us use the results for weak link to suggest a better estimation of $E_S$ in a homogeneous wire. There, the spatial extent of the phase-slip core is of the order of $\xi$ [1]. Let us separate the wire into pieces of the length $l_c$ and treat each piece as a diffusive weak link of corresponding resistance, $R_c = R'l_c$. We can find $l_c$ by comparing the critical current of a single weak link, $I_c = 1.32\,\pi\Delta/2eR_c$, and that of a homogeneous wire, $I_{cw} = \pi\Delta/3\sqrt{3}eR'\xi$ [25]. This gives $l_c \approx 3.43\,\xi$ and $E_S^* = 1.08\,\Delta(G_Q R'\xi)^{-1/2} e^{-0.360/G_Q R'\xi}$ per link. The amplitudes of the pieces add to $E_S = E_S^* L/l_c$.

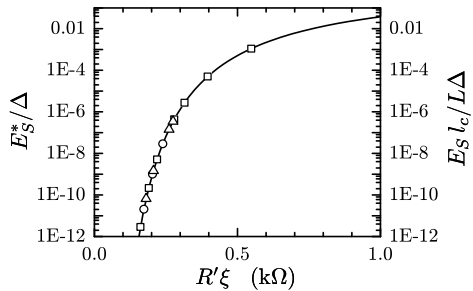

FIG. 4. Estimation of phase-slip amplitude for a long diffusive superconducting wire. Squares (□), circles (○), and triangles (△) give the estimated values of $R'\xi$ for superconducting nanowire samples as reported in [8], [6], [10], respectively.

The amplitude $E_S^*$ versus $R'\xi$ is plotted in Fig. 4 along with several values of $R'\xi$ for fabricated nanowires. Owing to exponential dependence on $R'\xi$, the phase-slip amplitude varies by nine orders of magnitude. We conclude that for most wires the expected $E_S^*$ is smaller than $10^{-6}\Delta$, with an exception of Ref. [8] where the wires have been fabricated by metal coating of a nanotube.

Let us use the above formula to estimate expected homogeneity of $E_S$ in realistic wires. We assume that fabrication imperfections induce normally distributed fluctuations of $G_c$ in each weak link with standard deviation $\delta G_c$. For $\delta G_c = 0$, the total $E_S$ scales with the length. However, if the fluctuations of $E_S^*$ are sufficiently large, the total $E_S$ can be just dominated by a single weak link of the lowest conductance. The criterion of crossover between these two regimes is derived to be $\ln(L/l_c) = (4.64\,\text{k}\Omega/R'\xi)^2(\delta G_c/G_c)^2$. It sharply depends on $R'\xi$. Let us assume $\delta G_c/G_c = 20\%$, a typical width variation of ultra-narrow wires. For smallest experimental $R'\xi$ in Fig. 4, the homogeneity is only realized if $L > 10^{17}\xi$! For the largest $R'\xi$, $L > 60\xi$ would suffice. The smallest possible $\delta G_c$ is determined by mesoscopic fluctuations. For the quantity given by Eq. (1), these fluctuations have been computed in [21]. Substitution leads to the homogeneity criterion $\ln(L/l_c) = (1/8)\ln(G_c/G_Q)$ [19]. This criterion is not restrictive for the values of $R'\xi$ in Fig. 4.

We see that even for apparently homogeneous wires $E_S$ may be strongly inhomogeneous. In addition, high values of $E_S$ are hard to achieve for the wires under experimental consideration. We suggest that fabrication of an *artificial* weak link may solve the problem. To do so, one can try to reduce selectively the wire width in a given point by, say, a factor of 2, either by laser or ion beam.

In conclusion, we have studied the quantum phase slips generated at a weak inhomogeneity in a superconducting wire. We have shown that the phase-slip action can be approximated by its topological part with accuracy better than 6%, thereby establishing a correspondingly accurate analytic estimate for the phase-slip amplitude. We have analyzed the consequences of that estimation when applied to realistic, imperfectly homogeneous wires. We suggest the fabrication of an artificial weak link would provide a better control needed for practical realization of the phase-slip devices.

The authors are indebted to J. E. Mooij for valuable discussions. This research was supported by the Dutch Science Foundation NWO/FOM.

# Supplementary material for "Quantum phase slips in superconducting wires with weak inhomogeneities"

Mihajlo Vanević and Yuli V. Nazarov

*Kavli Institute of Nanoscience, Delft University of Technology, 2628 CJ Delft, The Netherlands*

( Dated: February 13, 2012)

The Supplementary material is organized as follows. In Sec. 1, we obtain the action for a homogeneous superconducting wire which is modelled as an $LC$ transmission line. This action involves a time-dependent phase distribution $\phi(\tau, x)$ along the wire; after taking into account the boundary conditions for the phase at the weak link and at wire ends, the action of the wire is expressed in terms of a time-dependent phase drop $\phi(\tau)$ at the weak link only. This is the only dynamical variable in our model.

Section 2 is devoted to the action of the weak link. Here we obtain the action for a generic coherent weak link with arbitrary distribution of transmission eigenvalues. We demonstrate that in the stationary case, the obtained general expression for action in Eq. (A.5) reproduces the well-known Josephson current-phase relation of a generic contact, Eq. (A.7).

In Section 3, we study a *nonstationary* tunnel limit of Eq. (A.5) and recover the action for a superconducting tunnel junction, Eq. (A.8). We minimize this action and obtain the optimal $2\pi$ instanton. It turns out that this result obtained in the tunnel limit can be used for a general weak link as well.

The case of a general weak link is analyzed in Sec. 4. We show that the action in general consists of two parts: the topological part of the action which depends on the winding number only [Eq. (A.10)], and a non-topological part which depends on details of the time-dependent phase $\phi(\tau)$. The topological part represents a lower bound of the action. We find that it gives a good approximation to the action for optimal instanton. We argue that the optimal instanton for a general weak link is close to the one in the tunnel limit.

The calculation of the phase-slip amplitude $E_S$ is outlined in Sec. 5. We focus on the regime in which the phase slips are dominated by the weak link and do not depend on the properties of the wire (regions $I$, $II$ in Fig. 2 of the manuscript). The exponential factor in the phase-slip amplitude is to a good accuracy given by the topological part of the action. The pre-exponential factor in the amplitude is computed using the standard instanton technique. Importantly, this prefactor exhibits a divergence at large energies which can be cancelled by a renormalization of transmission eigenvalues. After renormalization, we obtain the amplitude $E_S$ for the phase-slips pinned at the weak link, Eq. (1) of the manuscript. This is the main result of the present work.

In Sec. 6, we use the obtained result for the weak link and suggest a better estimate of the phase-slip amplitude for homogeneous wires. We analyze the conductivity fluctuations in the wire due to fabrication imperfections and obtain the condition under which the wire can be considered homogeneous in the context of phase slips. We find that in general, the phase-slip amplitude $E_S$ has a large dispersion for the wires fabricated. We suggest it would be beneficial to fabricate a weak link artificially. Such a weak link would not destroy a phase-slip junction and would provide a better control for $E_S$.

### 1. Action of a superconducting wire

We model the superconducting wire as an $LC$ transmission line. The action of the wire reads

$$\mathcal{S}_\mathrm{w}[\phi(\omega,x)] = \frac{1}{8\pi e^2}\int d\omega \int_0^L dx \left(\frac{1}{2\mathcal{L}'(\omega)}\,|\partial_x\phi(\omega,x)|^2 \right.$$
$$\left. + \frac{C'\omega^2}{2}\,|\phi(\omega,x)|^2\right). \quad (A.1)$$

Let us assume the weak link is positioned at the point $x = L_1$. The boundary condition for the phase at the weak link is $\phi(\tau) = \phi(\tau, x = L_1 + 0) - \phi(\tau, x = L_1 - 0)$, where $\phi(\tau)$ is the phase drop at the link. The wire forms a superconducting loop and the phase at the wire ends satisfies $\phi(\tau, 0) = \phi(\tau, L)$. Taking into account the above boundary conditions for the phase at the weak link and at wire ends, the action $\mathcal{S}_\mathrm{w}$ can be expressed in terms of a time-dependent phase drop $\phi(\tau)$ at the link,

$$\mathcal{S}_\mathrm{w}[\phi] = \frac{1}{4\pi G_Q}\int_0^\infty \frac{d\omega}{2\pi}\,\omega\,Y(\omega)\,|\phi(\omega)|^2, \quad (A.2)$$

where the imaginary-frequency admittance $Y(\omega) = [\mathcal{L}'(\omega)/C']^{-1/2}[\tanh(\omega L_1/v_p) + \tanh(\omega L_2/v_p)]^{-1} - (L\mathcal{L}'_k\omega)^{-1}$, $L_1$ ($L_2$) is the length of the wire left (right) from the weak link, and $v_p(\omega) = 1/\sqrt{\mathcal{L}'(\omega)C'}$.

In the above formulas, $\mathcal{L}'(\omega)$ stands for imaginary-frequency inductance which is obtained by analytic continuation of the impedance [1], $\mathcal{L}'(\omega) = Z'_s(-i|\omega|)/|\omega|$. Here $Z'_s(\omega) = R'\sigma_n/\sigma(\omega)$ where $\sigma_n$ is the normal-state conductivity and $\sigma(\omega) = \sigma_1(\omega) - i\sigma_2(\omega)$ is the complex conductivity of a diffusive superconductor. At zero temperature, the real and imaginary parts $\sigma_{1,2}(\omega)$ are given by [2]

$$\frac{\sigma_1(\omega)}{\sigma_n} = \left(1 + \frac{2\Delta}{\omega}\right)E(k) - \frac{4\Delta}{\omega}K(k) \quad (A.3)$$

for $\omega > 2\Delta$ and $\sigma_1(\omega) = 0$ otherwise, and

$$\frac{\sigma_2(\omega)}{\sigma_n} = \frac{1}{2}\left(1 + \frac{2\Delta}{\omega}\right)E(k') - \frac{1}{2}\left(1 - \frac{2\Delta}{\omega}\right)K(k') \quad (A.4)$$

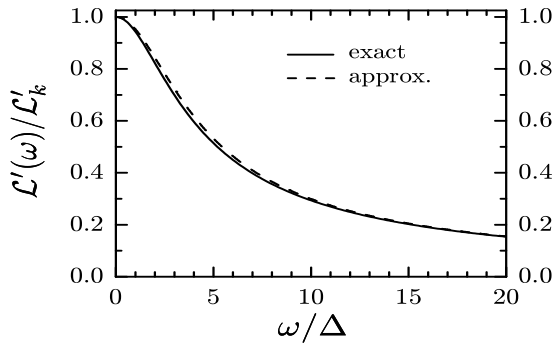

FIG. A1. Imaginary-frequency inductance $\mathcal{L}'(\omega)$ of a superconducting diffusive wire at zero temperature normalized to the kinetic inductance $\mathcal{L}'_k$: Exact $\mathcal{L}'(\omega)$ (solid) and approximation $\mathcal{L}'(\omega) \approx \mathcal{L}'_k \left[(\omega/\pi\Delta)^2 + 1\right]^{-1/2}$ (dashed line).

with $k = (\omega - 2\Delta)/(\omega + 2\Delta)$, $k' = \sqrt{1-k^2}$, and $K$, $E$ being complete elliptic integrals of the first and the second kind, respectively.

The analytic continuation $\sigma(-i|\omega|)$ is performed separately in domains $|\omega| < 2\Delta$ and $|\omega| > 2\Delta$. The resulting $\mathcal{L}'(\omega)$ is shown in Fig. A1. The inductance $\mathcal{L}'(\omega)$ is real and even function of frequency, continuous at $\omega = 2\Delta$. It takes into account the fact that the wire is inductive at subgap energies $\omega \ll 2\Delta$ with $\mathcal{L}'$ equal to kinetic inductance $\mathcal{L}' = \mathcal{L}'_k \equiv R'/\pi\Delta$, and resistive at large energies $\omega \gg 2\Delta$ with $\mathcal{L}'(\omega) = R'/\omega$. The inductance $\mathcal{L}'(\omega)$ can be approximated by $\mathcal{L}'(\omega) \approx \mathcal{L}'_k \left[(\omega/\pi\Delta)^2 + 1\right]^{-1/2}$ in the whole range of $\omega$ (cf. Fig. A1, dashed line).

### 2. Action of the weak link

The action $\mathcal{S}_c$ of a superconducting tunnel junction has been obtained in Refs. [3, 4]. We generalize the result to generic coherent weak link along the lines of Ref. [5]. The action reads

$$\mathcal{S}_c[\phi] = -\frac{1}{2}\sum_p \operatorname{Tr}\ln\left(1 + \frac{T_p}{4}(\{\hat{G}_1, \hat{G}_2\} - 2)\right), \quad (A.5)$$

where $\{T_p\}$ are spin-degenerate transmission eigenvalues. Equation (A.5) is valid for arbitrary set of $\{T_p\}$. The imaginary-time Green's functions $\hat{G}_j$ left and right from the contact are given by

$$\hat{G}_j(\tau, \tau') = e^{i\phi_j(\tau)\hat{\tau}_3/2}\,\hat{G}_0(\tau - \tau')\,e^{-i\phi_j(\tau')\hat{\tau}_3/2} \quad (A.6)$$

where $\phi_j(\tau)$ are the corresponding phases, $\hat{G}_0(\omega) = (\omega\hat{\tau}_3 + |\Delta|\hat{\tau}_1)/\sqrt{\omega^2 + |\Delta|^2}$ is the Green's function for a homogeneous superconductor, and $\hat{\tau}_i$ are Pauli matrices in electron-hole (Nambu) space. The product of Green's functions and the trace operation in Eq. (A.5) are understood in terms of convolutions over time (energy) and Nambu indices. We see that the action $\mathcal{S}_c$ depends on the phase difference $\phi(\tau) = \phi_2(\tau) - \phi_1(\tau)$ only.

Let us apply Eq. (A.5) in two limiting cases: (i) generic superconducting contact with a constant phase difference $\phi = \text{const}$ and (ii) superconducting tunnel junction ($T_p \ll 1$) with a fluctuating phase difference $\phi(\tau)$.

In the stationary case (i), the Josephson current is given by $I(\phi) = 2e\,\delta\mathcal{S}_c/\delta\phi$. The calculation of $\mathcal{S}_c$ simplifies considerably since the Green's functions $\hat{G}_j$ are diagonal in energy. We proceed with calculation of $I(\phi)$ using Eq. (A.5) in energy representation, where the trace over energy is understood in terms of summation over Matsubara frequencies, $\operatorname{Tr}_\omega = T\sum_{\omega_n}$ with $\omega_n = (2n+1)\pi T$ and $T$ being the temperature. We obtain

$$I(\phi) = \frac{e|\Delta|}{2}\sin(\phi)\sum_p \frac{T_p|\Delta|}{E_p}\tanh\left(\frac{E_p}{2T}\right) \quad (A.7)$$

where $E_p = |\Delta|\sqrt{1 - T_p\sin^2(\phi/2)}$. Equation (A.7) coincides with the generalized Josephson current-phase relation for the superconducting contact with arbitrary transmission eigenvalues [6].

### 3. Tunnel limit

As regards the nonstationary case, the calculation of the action in Eq. (A.5) is difficult because the Green's functions are no longer diagonal in energy. In general, to compute $\mathcal{S}_c$ for a time-dependent $\phi(\tau)$, the diagonalization of the operator $\{\hat{G}_1, \hat{G}_2\}$ is required. However, the action simplifies considerably in the tunnel limit (ii) and reduces to the trace of the products of the Green's functions, $\mathcal{S}_t = -(G_c/8G_Q)\operatorname{Tr}[\{\hat{G}_1, \hat{G}_2\} - 2]$. The trace can be computed in the original basis in which $\hat{G}_1$ and $\hat{G}_2$ are defined. We perform the calculation in time representation and obtain

$$\begin{aligned}\mathcal{S}_t[\phi] = {}& \frac{G_c}{G_Q}\frac{\Delta^2}{\pi^2}\int d\tau d\tau' \\ & \times \left[\sin^2\left(\frac{\phi(\tau) - \phi(\tau')}{4}\right) K_1^2(|\tau - \tau'|\Delta) \right.\\ & \left. + \sin^2\left(\frac{\phi(\tau) + \phi(\tau')}{4}\right) K_0^2(|\tau - \tau'|\Delta)\right]. \quad (A.8)\end{aligned}$$

Here $K_0$ and $K_1$ are the modified Bessel functions of the second kind which satisfy $K_0^2(y) \approx \ln^2(y)$ and $K_1^2(y) \approx 1/y^2$ for $y \ll 1$, and $K_0^2(y) \approx K_1^2(y) \approx \pi e^{-2y}/2y$ for $y \gg 1$. Equation (A.8) coincides with the action of a superconducting tunnel junction obtained in Refs. [3, 4].

We conclude this section with analysis of the action $\mathcal{S}_t$ which we will use later in a discussion of the general action $\mathcal{S}_c$. Let $\phi(\tau)$ be a time-dependent phase difference which changes from $\phi(-\infty) = 0$ to $\phi(\infty) = 2\pi$ during the time interval $\tau_p$. For the slow phase change $\tau_p \gg 1/\Delta$, the action $\mathcal{S}_t$ becomes local in time. It is dominated by the term proportional to $K_0^2$ and assumes the usual Josephson form, $\mathcal{S}_t = \int d\tau\, E_J(1-\cos\phi)$, where $E_J = G_c \Delta/4G_Q$. The action in this limit scales linearly with $\tau_p$, $\mathcal{S}_t \sim (G_c/G_Q)\tau_p\Delta \gg G_c/G_Q$. On the other hand, for the fast phase change $\tau_p \ll 1/\Delta$, the action $\mathcal{S}_t$ is dominated by the term proportional to $K_1^2$ and diverges logarithmically with vanishing $\tau_p$, $\mathcal{S}_t \sim (G_c/G_Q)\ln(1/\tau_p\Delta)$. As a result, the optimal instanton profile $\phi(\tau)$ which minimizes the action has duration $\tau_p \sim 1/\Delta$ and the minimal action is $\mathcal{S}_t \sim G_c/G_Q$.

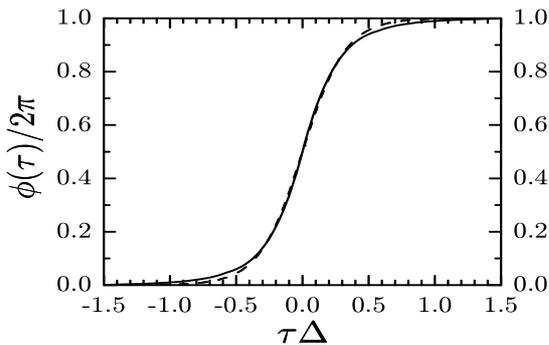

FIG. A2. The instanton profile $\phi(\tau)$ which minimizes $\mathcal{S}_t$ (solid line). The optimal one-parameter function $\phi_{\tau_p}(\tau)$ which minimizes $\mathcal{S}_t$ is shown for comparison (dashed line).

We minimize $\mathcal{S}_t$ in Eq. (A.8) by using the variational trial functions $\phi(\tau)$ which are constructed by the polynomial interpolation between the points $(\tau_i, \phi_i)$. Here $\tau_i$ are fixed and $\phi_i = \phi(\tau_i)$ are the variational parameters. The obtained optimal instanton profile $\phi(\tau)$ is shown in Fig. A2 (solid line) with the minimum of the action $\mathcal{S}_t = 0.528\, G_c/G_Q$. For comparison, we also perform minimization over the one-parameter family of variational functions $\phi_{\tau_p}(\tau) = \pi\,[\tanh(\tau/\tau_p) + 1]$. The minimum of the action in this case is found to be $0.535\, G_c/G_Q$ for $\tau_p\Delta = 0.325$. The optimal function $\phi_{\tau_p}(\tau)$ is shown in Fig. A2 (dashed line). We will see in the next section that the optimal instanton obtained here in the tunnel limit can also be used in the case of a general weak link.

### 4. General case: Topological and non-topological parts of the action

For an arbitrary transmission set $\{T_p\}$, the analytical calculation of the action in Eq. (A.5) is not feasible and we treat the problem numerically. However, the analysis of the numerical results enables us to formulate a good analytical approximation.

To start with, let us note that $\{\hat{G}_1, \hat{G}_2\} - 2 = -(\hat{G}_1 - \hat{G}_2)^2$, where we used the quasiclassical normalization condition of the Green's functions, $\hat{G}_j^2 = 1$. The action in Eq. (A.5) can be expressed in terms of the eigenvalues $\Lambda_n$ of the Hermitian operator $\hat{\Lambda} \equiv (\hat{G}_1 - \hat{G}_2)/2$,

$$\mathcal{S}_c[\phi] = -\frac{1}{2}\sum_{p,n}\ln(1 - T_p\Lambda_n^2). \quad (A.9)$$

We specify some properties of the eigenvalues $\Lambda_n$ which rely on the normalization condition of the Green's functions and do not depend on details of the time-dependent phase $\phi(\tau)$. First of all, since $\hat{G}_j$ are Hermitian and unitary, we find $|\Lambda_n| \leq 1$. In addition, operators $\hat{\Lambda}$ and $\hat{G}_1 + \hat{G}_2$ anticommute, $\{\hat{\Lambda}, \hat{G}_1 + \hat{G}_2\} = 0$. Let $\mathbf{v}$ be an eigenvector of $\hat{\Lambda}$, that is, $\hat{\Lambda}\mathbf{v} = \Lambda\mathbf{v}$. Then, if $(\hat{G}_1 + \hat{G}_2)\mathbf{v} \neq 0$, the vector $\mathbf{u} \equiv (\hat{G}_1 + \hat{G}_2)\mathbf{v}$ is the eigenvector of $\hat{\Lambda}$ which corresponds to the eigenvalue $-\Lambda$. This means that the eigenvalues of $\hat{\Lambda}$ typically appear in pairs $\pm\Lambda$ with the opposite sign.

In the special case $(\hat{G}_1 + \hat{G}_2)\mathbf{v} = 0$ for the certain eigenvectors $\mathbf{v}$ of $\hat{\Lambda}$. Thus, the corresponding eigenvalues $\Lambda$ do not come in pairs. These special eigenvectors satisfy $\hat{\Lambda}\mathbf{v} = \hat{G}_1\mathbf{v} = -\hat{G}_2\mathbf{v} = \Lambda\mathbf{v}$ and from $\hat{G}_j^2 = 1$ we find $|\Lambda| = 1$. Therefore, the special eigenvectors of $\hat{\Lambda}$ are the eigenvectors of *both* $\hat{G}_1$ *and* $\hat{G}_2$ with eigenvalues $\pm 1$. Depending on the sign of $\Lambda$, the special eigenvectors can have positive ($\Lambda = 1$) or negative ($\Lambda = -1$) *chirality*.

Let there be $N_+$ ($N_-$) special eigenvectors with positive (negative) chirality. Since all other eigenvalues come in pairs with the opposite sign, we find that $\text{Tr}(\hat{\Lambda}) = N_+ - N_-$. The trace is a topological property of $\hat{\Lambda}$ with respect to variations of $\phi(\tau)$. It is set by the winding number of the phase $N = \lfloor |\int_{-\infty}^{\infty} \dot{\phi}(\tau)\,d\tau|/2\pi \rfloor$, where $\lfloor \cdot \rfloor$ denotes the integer part. From Eq. (A.9) we see that the contribution of the special eigenvalues to the action is proportional to $N_+ + N_-$, where the difference $N_+ - N_-$ is constrained by the phase winding number $N = |N_+ - N_-|$. Since $N_+ + N_- \geq N$, the minimal contribution to the action is given by

$$\mathcal{S}_{c1}(N) = -\frac{N}{2}\sum_p \ln(1 - T_p) \quad (A.10)$$

and is achieved if the special eigenvectors of only one chirality are present. The sign of the chirality is determined by the direction (clockwise or counterclockwise) of the phase winding. The analysis presented here is analogous to the one carried out in Ref. [7].

The special eigenvalues give the topological contribution to the action $\mathcal{S}_{c1}$ which represents a lower bound for $\mathcal{S}_c$ and does not depend on details of time-dependent $\phi(\tau)$. In contrast, paired eigenvalues $\pm\Lambda$ are sensitive to time dependence of $\phi(\tau)$ and determine the magnitude of the action for the phase winding number fixed. For a general $\phi(\tau)$, the paired eigenvalues with $|\Lambda| \approx 1$ can occur [7]. Nonetheless, such paired eigenvalues would give a large contribution to the action on the order of $\mathcal{S}_{c1}$ and we expect them to be absent for *optimal* instanton $\phi(\tau)$ which minimizes the action.

The topological lower bound $\mathcal{S}_{c1}$ can be realized if there is an instanton $\phi(\tau)$ for which all paired eigenvalues $\Lambda_n$ are zero. In the normal state such instantons indeed exist and can even be found analytically [8]. This is not the case for superconducting action. However, one can expect the paired eigenvalues are small for *optimal* in-stanton $\phi(\tau)$ which minimizes the action. In that case the topological lower bound $\mathcal{S}_{c1}$ is a good approximation to the true action $\mathcal{S}_c$.

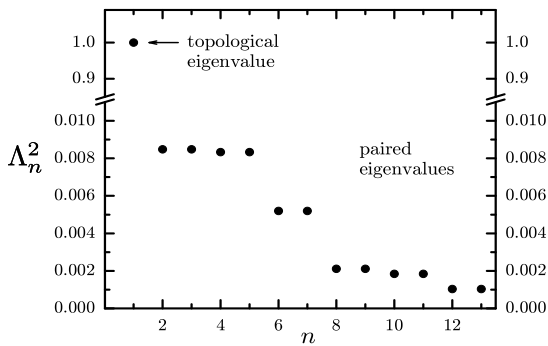

FIG. A3. The eigenvalues squared $\Lambda_n^2$ of $\hat{\Lambda} = (\hat{G}_1 - \hat{G}_2)/2$ for the $2\pi$ instanton in the tunnel limit. The main contribution to the action in Eq. (A.9) is topological, arising from $\Lambda = 1$.

We verify that this is indeed the case by computing the spectrum of $\hat{\Lambda}$ for the optimal $2\pi$ instanton (winding number $N = 1$) in the tunnel limit. This instanton has been obtained in the previous section. The action $\mathcal{S}_c$ depends on eigenvalues squared $\Lambda_n^2$ which are shown in Fig. A3. The main contribution to the action is topological $\mathcal{S}_{c1} = 0.5 \, G_c/G_Q$, arising from the special eigenvalue $\Lambda = 1$. We find that paired eigenvalues $\Lambda_n^2 \ll 1$ give a small correction $\mathcal{S}_{c2} = 0.028 \, G_c/G_Q$ to it of a few percent only. Formally, the exponential dependence of the phase-slip amplitude $E_S$ could amplify even this small error by orders of magnitude; yet, as shown in Fig. 3 of the manuscript, this does not happen for any $E_S$ of interest.

We argue that the optimal instanton for a general weak link is close to the one obtained in Sec. 3 in the tunnel limit. For optimal $\phi(\tau)$, the paired eigenvalues are expected to be small and their contribution in Eq. (A.9) can be expanded in series of $T_p\Lambda_n^2 \ll 1$. This expansion formally coincides with the one in the tunnel limit, $T_p \ll 1$. As regards the topological part of the action, it is given by the winding number only and is independent on the details of $\phi(\tau)$. Therefore, to first order in $\Lambda_n^2 \ll 1$, the optimal instanton is universal: It is the same regardless the type of the weak link and can be determined from the minimization of the action $\mathcal{S}_t$ in the tunnel limit (see Sec. 3).

The universality of the instanton solution $\phi(\tau)$ implies the leading correction to the topological part of the action is independent on the details of transmission distribution and equals $0.03 \, G_c/G_Q$. This means the topological part is a good approximation to the action also beyond the tunnel limit. Indeed, comparing the topological contribution and the leading correction, we observe that the accuracy of 6% that is achieved in the tunneling limit is in fact the lowest one. This is because the topological contribution grows diverging for a perfect transmission, while the leading correction remains the same. The relative accuracy beyond tunnel limit is thus improved owing to increased topological part of the action. For example, in case of a diffusive weak link that is characterized by transmission distribution $\rho(T) = (G_c/G_Q)(2T\sqrt{1-T})^{-1}$ [13], the instanton action is $\mathcal{S}_c \approx 1.26 \, G_c/G_Q$ while its topological part $\mathcal{S}_{c1} = (\pi^2/8) \, G_c/G_Q \approx 1.23 \, G_c/G_Q$.

### 5. Pase-slip amplitude for the weak link

The phase-slip amplitude $E_S$ gives a qubit level splitting due to instantons connecting energy degenerate minima [9]. The pre-exponential factor in the amplitude $E_S$ is obtained by utilizing the quadratic expansion of the action with respect to the instanton trajectory. The instanton approach is justified when the instanton action is big. This is equivalent to the condition that the instanton amplitude $E_S$ is much smaller than a typical inverse time-scale of the action $\Delta$.

The phase-slip amplitude is given by [10]

$$E_S = 2 \left( \frac{1}{2\pi} \int d\tau \, \dot{\phi}_{\rm in}^2(\tau) \right)^{1/2} (D')^{-1/2} \, e^{-\mathcal{S}_{\rm in}}, \quad (A.11)$$

where $\mathcal{S}_{\rm in} = \mathcal{S}_c[\phi_{\rm in}]$ and $\phi_{\rm in}(\tau)$ is the instanton solution which minimizes the action and satisfies the boundary conditions $\phi_{\rm in}(-\infty) = 0$ and $\phi_{\rm in}(\infty) = 2\pi$ (0 and $2\pi$ being the stationary minima of the action). The above expression for $E_S$ is valid for arbitrary effective action, including the nonlocal ones [11]. In this situation $\int d\tau \, \dot{\phi}_{\rm in}^2(\tau)$ is no longer proportional to $\mathcal{S}_{\rm in}$ (as it would be the case in a potential field) and has to be calculated explicitly from the instanton solution.

The ratio of functional determinants

$$D' = \frac{{\det}'(H_1)}{\det(H_0)} \quad (A.12)$$

in Eq. (A.11) describes the fluctuations $\chi(\tau)$ with respect to trivial $[\phi(\tau) = \chi(\tau)]$ and instanton $[\phi(\tau) = \phi_{\rm in}(\tau) + \chi(\tau)]$ trajectories. Here, the kernels $H_0$ and $H_1$ are defined by

$$H_0(\tau,\tau') = \left.\frac{\delta^2 \mathcal{S}_c}{\delta\chi(\tau)\delta\chi(\tau')}\right|_{\phi=0},$$
$$H_1(\tau,\tau') = \left.\frac{\delta^2 \mathcal{S}_c}{\delta\chi(\tau)\delta\chi(\tau')}\right|_{\phi=\phi_{\rm in}(\tau)}. \quad (A.13)$$

The prime $'$ in Eq. (A.12) denotes that the zero eigenvalue intrinsic to instanton is omitted in the numerator.

In the following we outline the method of calculation of $D'$. We first note that the kernel $H_0(\tau,\tau') = H_0(\tau-\tau')$ depends on time difference only and is diagonal in energy, $H_0(\omega) = (2E_J/\pi) \, E(i\omega/2\Delta)$. Here $E_J = G_c\Delta/4G_Q$ and $E(k)$ is the complete elliptic integral of the first kind. At large frequencies $|\omega| \gg \Delta$ the kernel $H_0$ is linear in $\omega$,

$$H_0(\omega) = E_J|\omega|/\pi\Delta, \quad (A.14)$$

while at zero frequency $H_0(\omega = 0) = E_J$. After we single out $E_J$ which is the lowest eigenvalue of $H_0$, Eq. (A.12) becomes

$$D' E_J = \frac{\det'(H_1)}{\det'(H_0)}. \tag{A.15}$$

The prime $'$ in the denominator denotes that the lowest eigenvalue of $H_0$ is omitted. The product $D' E_J$ is dimensionless with both numerator and denominator on the right-hand side of Eq. (A.15) having equal number of eigenvalues.

Since $H_0$ is already diagonal in energy, the calculation of $D'$ reduces to diagonalization of $H_1$. We proceed as follows. We compute the low-lying eigenvalues $h_m^{(i)}$ of $H_i$ and the ratio $q_m = h_m^{(1)}/h_m^{(0)}$ by imposing the hard-wall boundary conditions at large times $\tau_0$. The boundary conditions lead to energy discretization $\omega_m = m\delta\omega$, where $\delta\omega = \pi/\tau_0$ and $m = 1, 2, \ldots$. On the other hand, the eigenvalues of $H_1 \equiv H_0 + \delta H$ for large $m$ can be calculated perturbatively with respect to $H_0$. We find that the perturbative correction $(\delta H)_{mm}$ for large $m$ is constant: $\delta H = -(G_c/G_Q)(\sum_n \Lambda_n^2)/2\tau_0$. After taking the logarithm of Eq. (A.15) and replacing the summation over $m$ by integration, we obtain

$$\ln(D' E_J) = \sum_{m=2}^{\omega_c/\delta\omega} \ln(q_m) - \left(\frac{G_c}{G_Q} \sum_n \Lambda_n^2\right) \frac{1}{4\pi} \int_{\omega_c}^{\infty} \frac{d\omega}{H_0(\omega)}. \tag{A.16}$$

Here, the frequency $\omega_c$ separates the contribution of low-lying eigenvalues (first term) and the contribution of large eigenvalues (second term on the right-hand side). The precise value of $\omega_c$ is not essential: once $\omega_c$ is large enough to justify the perturbative approach, all relevant low-lying eigenvalues are taken into account by the first term in Eq. (A.16) and the value of $D'$ no longer depends on the choice of $\omega_c$.

We focus on the last term in Eq. (A.16) which diverges logarithmically at large frequencies. This is related to the frequency dependence of the integral kernels in the action, Eq. (A.14). In principle, account of the wire capacitance might provide an upper cut-off $E_c$ needed. However, we find it more consistent to cancel the divergence by taking into account the renormalization of transmission eigenvalues. Indeed, it is known that Coulomb interaction leads to energy-dependent renormalization of $T_p$ [12]. From Eq. (A.16) we find $D' \propto (G_c\Delta/G_Q)^{-1}(\Delta/E_c)^{\sum_n \Lambda_n^2}$ and upon substitution in Eq. (A.11) we obtain for the amplitude

$$E_S \propto (E_c/\Delta)^{\sum_n \Lambda_n^2/2} e^{-S_{\rm in}}. \tag{A.17}$$

The action $S_{\rm in}$ is given by Eq. (A.9) with $\Lambda_n$ computed for instanton $\phi_{\rm in}(\tau)$. The renormalization of transmission eigenvalues reads

$$\frac{dT_p}{d\ln E} = \frac{T_p(1-T_p)}{\sum_p T_p}. \tag{A.18}$$

Correcting the transmissions in the action with the above equation cancels the ultraviolet divergence in $E_S$. It implies that $T_p$ in all formulas must be taken as measured experimentally, that is, at $E \simeq \Delta$ rather than at unphysically high energy.

This brings us to the expression for the amplitude

$$E_S = a \left(\frac{\Delta}{2\pi} \int d\tau\, \dot\phi_{\rm in}^2(\tau)\right)^{1/2} \left(\sum_p T_p\right)^{1/2} e^{-S_{\rm in}}, \tag{A.19}$$

where the prefactor $a$ is a constant given by the low-lying eigenvalues of the integral kernels [first term in Eq. (A.16)]. We compute $a \approx 0.8$ in the tunnel limit and $\int d\tau \dot\phi_{\rm in}^2 \approx 40.5\,\Delta$ for the instanton shown in Fig. A2. Taking into account only major topological contribution to the action $S_{c1} = (-1/2)\sum_p \ln(1-T_p)$, we recover $E_S$ given by Eq. (1) of the manuscript,

$$E_S \approx 2\Delta \sqrt{\sum_p T_p} \prod_p \sqrt{1-T_p}. \tag{A.20}$$

To make concrete predictions we have to specify the type of the weak link. For a diffusive weak link, the topological action $S_{c1} = (\pi^2/8)\, G_c/G_Q \approx 1.23\, G_c/G_Q$ (see Sec. 4) and the phase-slip amplitude $E_S \approx 2\Delta\sqrt{G_c/G_Q}\, e^{-1.23\, G_c/G_Q}$. The phase-slip amplitude is shown in Fig. 3 of the manuscript as a function of the weak-link resistance.

### 6. Pase-slip amplitude for a homogeneous wire

The analysis of the phase-slip amplitude carried out in the previous section pertains to a weak link which is longer than the mean free path but shorter than the superconducting coherence length $\xi$. Let us use the obtained results for weak link to suggest a better estimate of $E_S$ in a homogeneous wire. There, the spatial extension of the phase-slip core is on the order of $\xi$ [14]. Let us separate the wire into pieces of length $l_c$ and treat each piece as a diffusive weak link of the corresponding resistance $R_c = R' l_c$. We can find $l_c$ by comparing the critical current of a single weak link, $I_c = 1.32\,\pi\Delta/2eR_c$, and that of a homogeneous diffusive wire, $I_{\rm cw} = \pi\Delta/3\sqrt{3}eR'\xi$ [2]. This gives $l_c \approx 3.43\,\xi$ and the phase-slip amplitude

$$E_S^* \approx \frac{1.08\,\Delta}{\sqrt{G_Q R'\xi}}\, e^{-0.360/G_Q R'\xi} \tag{A.21}$$

per link. The amplitudes of the pieces add to the phase-slip amplitude of the wire,

$$E_S = E_S^* L/l_c. \tag{A.22}$$

The phase-slip amplitude $E_S^*$ versus $R'\xi$ is plotted in Fig. 4 of the manuscript along with several values of $R'\xi$ for fabricated nanowires. Because of exponential dependence on $R'\xi$, the phase-slip amplitude varies by nine orders of magnitude. We find that for most wires the expected $E_S^*$ is smaller than $10^{-6}\Delta$ except possibly for wires reported in [15].

Next, let us use Eq. (A.21) and discuss the conditions under which the wire can be considered homogeneous with respect to quantum phase slips. We assume that fabrication imperfections introduce normally distributed fluctuations of $G_c$ in each weak link with standard deviation $\delta G_c$. This amounts to the average amplitude per link $\langle E_S^* \rangle = \tilde{E}_S^* e^{(\alpha \delta G_c/G_Q)^2/2}$ with variance $\mathrm{Var}(E_S^*) = (\tilde{E}_S^*)^2 (e^{2(\alpha \delta G_c/G_Q)^2} - e^{(\alpha \delta G_c/G_Q)^2})$. Here $\alpha = \pi^2/8$ for a diffusive weak link.

For ideally homogeneous wire $\delta G_c = 0$ and the total $E_S$ is proportional to the length, Eq. (A.22). However, if the fluctuations of $E_S^*$ per link are sufficiently large, the total $E_S$ can be just dominated by a single weak link of the lowest conductance. The criterion under which the wire can be considered homogeneous with respect to phase-slips is that the average amplitude given by $\langle E_S \rangle = (L/l_c)\langle E_S^* \rangle$ is larger than its standard deviation, $\delta E_S \equiv \sqrt{\mathrm{Var}(E_S)}$. We assume the imperfections along the wire are uncorrelated. This gives $\mathrm{Var}(E_S) = (L/l_c)\mathrm{Var}(E_S^*)$ and the homogeneity condition $\langle E_S \rangle > \delta E_S$ becomes $L/l_c > e^{(\alpha \delta G_c/G_Q)^2} - 1$. Using $l_c = 3.43\,\xi$ and $\alpha = \pi^2/8$ we obtain the wire can be considered homogeneous if it is longer than

$$\ln(L/l_c) > (4.64\,\mathrm{k\Omega}/R'\xi)^2 \,(\delta G_c/G_c)^2. \qquad (A.23)$$

The above criterion sharply depends on $R'\xi$. Let us assume $\delta G_c/G_c = 20\%$ which is the typical width variation of ultra-narrow wires [16, 17]. For smallest experimental value $R'\xi \approx 0.15\,\mathrm{k\Omega}$ shown in Fig. 4 of the manuscript, the homogeneity is only realized if $L > 10^{17}\xi$! For the largest $R'\xi \approx 0.55\,\mathrm{k\Omega}$, the wire length $L > 60\,\xi$ would be sufficient.

The criterion derived in Eq. (A.23) takes into account fluctuations $\delta G_c$ due to fabrication or material imperfections. On the other hand, the smallest possible $\delta G_c$ is given by mesoscopic fluctuations. For the quantity $\mathcal{S}_{c1} = (-1/2)\sum_p \ln(1-T_p)$ ($p$ labels spin-degenerate transport channels) these fluctuations were computed in [8] and give $(\delta \mathcal{S}_{c1})^2 \equiv \mathrm{Var}(\mathcal{S}_{c1}) = (N_{\mathrm{dc}}/16)\ln(G_c/G_Q)$. Here, $N_{\mathrm{dc}} = 1-8$ is the total number of massless cooperon/diffuson modes; $N_{\mathrm{dc}} = 2$ is consistent with superconductivity and weak spin-orbit interaction. For the phase-slip amplitude per weak link $E_S^* = 2\Delta\sqrt{G_c/G_Q}\,e^{-\mathcal{S}_{c1}}$ we obtain $\langle E_S^* \rangle = \tilde{E}_S^* e^{(\delta \mathcal{S}_{c1})^2/2}$ and $\mathrm{Var}(E_S^*) = (\tilde{E}_S^*)^2 (e^{2(\delta \mathcal{S}_{c1})^2} - e^{(\delta \mathcal{S}_{c1})^2})$. Similarly as before, the wire can be considered homogeneous if the total phase-slip amplitude satisfies $\langle E_S \rangle > \delta E_S$, which reduces to $\ln(L/l_c) > (1/8)\ln(G_c/G_Q)$. This criterion only weakly depends on $R'\xi$ and is not restrictive for experimental samples shown in Fig. 4 of the manuscript.

We conclude that the total phase-slip amplitude $E_S$ typically has a large dispersion in the wires under experimental consideration. The fabrication of an artificial weak link will not destroy the phase-slip junction and can provide a better control for $E_S$.

---